\documentclass[aps,prl,preprint,nopacs,superscriptaddress]{revtex4}
\usepackage{graphicx}
\usepackage{verbatim}

\begin{document}

\title{Observation of topological order in a superconducting doped topological insulator}

\author{L. Andrew Wray}
\affiliation{Department of Physics, Joseph Henry Laboratories, Princeton University, Princeton, NJ 08544, USA}
\affiliation{Advanced Light Source, Lawrence Berkeley National Laboratory, Berkeley, California 94305, USA}
\author{Suyang Xu}
\author{Yuqi Xia}
\affiliation{Department of Physics, Joseph Henry Laboratories, Princeton University, Princeton, NJ 08544, USA}
\author{Dong Qian}
\affiliation{Department of Physics, Joseph Henry Laboratories, Princeton University, Princeton, NJ 08544, USA}
\affiliation{Department of Physics, Shanghai Jiao-Tong University, Shanghai 200030, People's Republic of China}
\author{Alexei V. Fedorov}
\affiliation{Advanced Light Source, Lawrence Berkeley National Laboratory, Berkeley, California 94305, USA}
\author{Hsin Lin}
\author{Arun Bansil}
\affiliation{Department of Physics, Northeastern University, Boston, MA 02115, USA}
\author{Yew San Hor}
\author{Robert J. Cava}
\affiliation{Department of Chemistry, Princeton University, Princeton, NJ 08544, USA}
\author{M. Zahid Hasan}
\affiliation{Department of Physics, Joseph Henry Laboratories, Princeton University, Princeton, NJ 08544, USA}



\date{\today}

\maketitle

\textbf{Topological insulators embody a new state of matter characterized entirely by the topological invariants of the bulk electronic structure rather than any form of spontaneously broken symmetry \cite{Intro,TI_RMP,TIbasic,DavidNat1,DavidScience,DavidTunable,MatthewNatPhys,ChenBiTe}. Unlike the quantum Hall or quantum spin-Hall-like systems, the three dimensional topological insulators can host magnetism and superconductivity which has generated widespread research activity in condensed-matter and materials-physics communities.
Thus there is an explosion of interest in understanding the rich interplay between topological and the broken-symmetry states (such as superconductivity)  \cite{FuSCproximity,FuHexagonal,FuNew,QiTopoSC,MudrySurfSC,NagaosaUnconventional,PairStr,AshvinChiral,ImpurityStates,FerroSplitting, KitaevClass, SchnyderClass}, greatly spurred by proposals that superconductivity introduced into certain band structures will host exotic quasiparticles which are of great interest in quantum information science. The observations of superconductivity in doped Bi$_2$Se$_3$ (Cu$_x$Bi$_2$Se$_3$) and doped Bi$_2$Te$_3$ (Pd$_x$-Bi$_2$Te$_3$ T$_c$ $\sim$ 5K) have raised many intriguing questions about the spin-orbit physics of these ternary complexes while any rigorous theory of superconductivity remains elusive \cite{HorSC,HorAllSC}. Here we present key measurements of electron dynamics in systematically tunable normal state of Cu$_x$Bi$_2$Se$_3$ (x=0 to 12$\%$) gaining insights into its spin-orbit behavior and the topological nature of the surface where superconductivity takes place at low temperatures. Our data reveal that superconductivity occurs (in samples compositions) with electrons in a bulk relativistic kinematic regime and we identify that an unconventional doping mechanism causes the topological surface character of the undoped compound to be preserved at the Fermi level of the superconducting compound, where Cooper pairing occurs at low temperatures. These experimental observations provide important clues for developing a theory of superconductivity in highly topical Bi-based 3D topological insulators.}

Bismuth selenide has been experimentally discovered by angle-resolved photoemission spectroscopy (ARPES) to be a topological insulator with a large bulk bandgap \cite{Intro, DavidTunable,MatthewNatPhys}. Spin-resolved photoemission studies reveal that surface electrons in Bi$_2$Se$_3$ form a Dirac cone spanning the bulk insulating gap, composed of spin-momentum-locked helical states \cite{DavidTunable,MatthewNatPhys}.
Recent theory suggests that just as topological insulators can be identified from band structure alone, the topological properties of spin-orbit superconductors can be determined from the connectivity of bulk and surface band structure \cite{MudrySurfSC, NagaosaUnconventional, FuNew, QiTopoSC}. Superconducting pairing induced on a two dimensional topological surface state via proximity to a bulk superconductor was shown to be inherently unconventional, with the potential to host non-Abelian vortices \cite{FuSCproximity} and defying traditional classification as a singlet or triplet form of Cooper pairing \cite{FuNew}. To ensure that a material is capable of hosting two dimensional vortices in the superconducting state, the spin-polarized surface states are must be non-degenerate with the bulk bands where Cooper pairing takes place. It has been found that 10$\%$ copper is needed to bring about superconductivity in bulk Cu$_x$Bi$_2$Se$_3$, with a relatively large transition temperature of T$_C$=3.8$^o$ K confirmed by the observation of the Meissner effect \cite{HorSC}. All known measured band structures as well as band calculations of undoped Bi$_2$Se$_3$ suggest that the topological surface states cannot remain non-degenerate at the Fermi level under the carrier doping that is required to induce superconductivity in Cu$_x$Bi$_2$Se$_3$. Therefore it was expected that the topological effects on the surface of Cu$_x$Bi$_2$Se$_3$ would be washed out at the superconducting compositions (comparisons in online Supplementary Information).

In clear contrast, as we show here a direct band structure measurement of superconducting Cu$_{0.12}$Bi$_2$Se$_3$ reveals that the surface states are distinctly isolated from the bulk conduction bands where bulk superconductivity takes place (see Fig. 3). Motivated by this puzzling observation we have systematically investigated the doping process by examining several stoichiometric crystals with copper added (Cu$_x$Bi$_2$Se$_3$, x=0, 0.01, 0.05, 0.12) and contrasting these doping levels with a case in which copper is substituted for bismuth as Cu$_{0.1}$Bi$_{1.9}$Se$_3$. These results, summarized in Fig. 1c, suggest  that copper atoms can add holes or electrons depending on their net preference for occupying different kinds of site in the Bi$_2$Se$_3$ lattice. Copper atoms intercalated between van der Waals-like bonded selenium planes (Cu$_{int}$) in the crystal are thought to be single electron donors, while substitutional defects in which copper replaces bismuth in the lattice (Cu$_{Bi}$) contribute two holes to the system which is similar to the behavior reported in the impurity atom study on Bi$_2$Se$_3$ matrix\cite{CuAmphoteric}. The qualitative result of copper addition is a gradual enlargement of the observed Fermi surface from electron doping and, further surprisingly, a strong reduction in surface electron velocities through reshaping of the surface state conduction band. The slope of the surface state band represents particle velocity, and is globally reduced in amplitude by 30$\%$ above the Dirac point when copper doping of x=0.05 is added to the stoichiometric compound (see Fig. 2c). Assuming that the surface state Dirac point energy is fixed relative to bulk bands, reducing the slope of the surface state increases separation between the bulk and surface conduction bands, a characteristic that increases stability of the topological surface state and is of great importance for stabilizing the topological nature of superconductivity. This effect acts together with the nonlinear doping mechanism reported here to preserve the two dimensional topological surface state at superconducting doping. Therefore our results suggest that the crystal surface of Cu$_x$Bi$_2$Se$_3$ can host two dimensional vortices, which is a prerequisite for observing non-Abelian statistics on the surface.

Addition of copper causes the surface state kinetics to become strongly hexagonally anisotropic, (Fig. 2b) a deformation that makes the surface susceptible to spin-fluctuation or magnetic instabilities \cite{FuHexagonal}. This is due to the fact that the topological surface states in Bi$_2$Se$_3$ are spin polarized \cite{DavidTunable}. Carrier density in the surface state is much greater than density in the bulk, as estimated by the Luttinger count applied to the photoemission (ARPES) data (Fig. 1c), suggesting that the surface carries a screened negative charge within the normal state of the superconductor. We also observed that attempting to force the creation of Cu$_{Bi}$ replacement defects by adding less bismuth results in very weak hole doping for Cu$_{0.1}$Bi$_{1.9}$Se$_3$ (Fig. 1c), raising the bulk conduction band entirely above the Fermi level. Upon doping into the superconducting regime (x=0.12), the Fermi energy is found to be raised to 0.25 eV above the bulk conduction band minimum, placing the Fermi level in the linear (relativistic) regime of bulk band dispersion. Contrasting the bulk band lineshape with classical and relativistic Dirac-like energy dispersions with the same mass (M=0.155 M$_e$) reveals that electron kinetics begin to enter the linear relativistic regime within $\sim$0.1 eV of the Fermi level (Fig. 3f). A slight bend in the dispersion centered near 90 meV binding energy (Fig. 3f inset) may suggest electron-boson interactions in the system, also ubiquitously observed in other superconducting materials. Therefore, Cooper pairing in the superconducting state takes place in this relativistic regime where the chemical potential lies. A three dimensional massive Dirac-like dispersion in topological spin-orbit materials is expected as a direct result of the band inversion mechanism that causes the topological insulator state, and has also been experimentally observed in other topological insulator Bi$_{1-x}$Sb$_x$ alloys \cite{DavidNat1}. In Cu$_x$Bi$_2$Se$_3$, calculations predict that band inversion occurs at the $\Gamma$-point (k$_x$=k$_y$=k$_z$=0) in the center of the bulk conduction band leading to a spin-orbit induced Dirac-like bulk band in clear qualitative agreement with our experiments.

Several features of this unusual spin-orbit band structure provide critical insights into characteristics of the superconducting wavefunction. Bulk Fermi momenta of 0.110$\pm$3 $\AA^{-1}$ and 0.106$\pm$3 $\AA^{-1}$ are observed along the $\overline{\Gamma}$-$\overline{M}$ and $\overline{\Gamma}$-$\overline{K}$ directions respectively. Varying incident energy to observe dispersion along the $\hat{z}$ axis ($\Gamma$-Z direction) reveals a Fermi momentum of 0.12$\pm$1 $\AA^{-1}$, suggesting that the bulk electron kinetics are three dimensionally isotropic. Carefully tracing the band (Fig. 3e,f) reveals a Fermi velocity of 3.5$\pm$2 eV$\times\AA$ along $\overline{\Gamma}$-$\overline{M}$ and 4.1$\pm$2 eV$\times\AA$ along $\overline{\Gamma}$-$\overline{K}$, estimated within 50 meV of the Fermi level. The gap between bulk valence and conduction bands appears to be unchanged upon copper doping. In a material specific band-structure based modeling of superconductivity, considering a competition between inter-site and intra-site orbital coupling suggests that the superconducting pairing symmetry in this system is partially determined by a ratio between the Dirac rest mass and doped chemical potential (``$\mu$") \cite{FuNew}, when the conduction band is fitted to a standard relativistic Hamiltonian:

\begin{equation}
H_0(k) = M^*\Gamma_0 + v_C^*(k_x\Gamma_1 + k_y\Gamma_2 + k_z\Gamma_3),
\end{equation}

where M$^*$ is the rest mass, v$_C$ is the velocity of the 3D Dirac band, and the $\Gamma_i$'s (i = 0; ... ; 3) are the 4$\times$4 Dirac Gamma matrices. For this discussion we assume that the v$_C$ is isotropic, although our data suggest that it may be smaller in the z-axis direction. The chemical potential of this Hamiltonian is equal to the sum of the rest mass and the binding energy of the valence band minimum, which can be experimentally evaluated from our photoemission data. In an ideal Dirac band structure, the rest mass will be equal to half the band gap, or 150 meV (giving M*=.15eV, $\mu$=0.15 eV + 0.25 eV, and $\frac{M^*}{\mu}\sim\frac{ 1}{3}$).  Due to asymmetries between conduction and valence band dispersion in the real experimental system, a larger effective rest mass is used to fit the curvature of the bottom of the conduction band in Fig. 3f, consistent with a ratio of $\frac{M^*}{\mu}$ =$\frac{ 3}{4}$.  These ratios are small enough to allow the rare theoretical possibility that the bulk superconducting wavefunction may have an odd (-1) parity symmetry value \cite{FuNew}. A dispersion anomaly that we have noted in the inset to Fig. 3f also effects low energy electron kinetics, making the bands more linear locally at the Fermi level. Dispersion anomalies, often referred to as ``kinks" in photoemission literature, are a common property of band structure in real superconducting materials \cite{SCproperties}. The presence of the kink depresses the value of $\frac{m}{\mu}$ in the favorable direction for realizing odd parity pairing proposed in Ref. \cite{FuNew}.

The observed spin-orbit band structure kinematics fixes several key parameters related to the superconductivity. The superconducting coherence length can be estimated from the average Fermi velocity we observe here to be about $2000\AA$ ($\xi_0$$\sim0.2\times\hbar v_F/K_BT_C=0.2\times3.8eV\AA/(K_B\times3.8^oK)=2000\AA$), assuming minimal scattering and a superconducting gap related to T$_C$ by the BCS formula. This value is about 1-2 orders of magnitude greater than that seen in the cuprates ($\xi_0$$\sim100-200\AA$) or cobaltates ($\xi_0$$\sim200\AA$) \cite{SCproperties,SCDong}. The phase ordering temperature scale, also estimated from our data, is approximately T$_\theta$ = 60,000$^o$K, four orders of magnitude larger than the superconducting critical temperature of 3.8$^o$K (T$_\theta$= $\xi_0\times\hbar^2n_e/2m^*$= 2000$\AA\times\hbar^2\times10^{20} cm^{-3}/(2\times.155 Me)$ = 60,000$^o$K). A large coherence length and high phase ordering temperature suggest that key properties of the pairing state can be described within a mean field picture. Based on this, we expect the superconducting gap to be about $\sim$0.6 meV (3.5$\times$K$_B$T$_C$/2=0.6 meV). This is much smaller than the state-of-the-art resolution of ARPES.

The fact that superconductivity occurs in the relativistically linear bulk bands in Cu$_x$Bi$_2$Se$_3$ as seen in our data allows unusual combinations of spin and orbital mixing in the bulk superconducting state, making pairing take an unusual form; more remarkably however, probing near the Fermi level, our data show that the spin-textured surface and doubly degenerate bulk bands are well separated by 0.04 $\AA^{-1}$ in momentum and an energy spacing of about $\Delta_E$=130 meV, establishing that bulk superconducting pairing occurs in the presence of a non-degenerate, spin polarized two dimensional topological surface state, and that theoretical constraints ensuring unconventional superconductivity induced in the topological surface band structure will define key properties of the pairing wavefunction \cite{TIbasic,FuSCproximity}. The remarkable preservation of the surface state occurs due to nonlinear doping and unusual distortions of the surface state that we have discovered here (Fig. 2c). This experimental observation is in contrast to first principles predictions, which suggest that the surface state will become degenerate and thus lose the protected spin-texture at the Fermi level due to intersection with the bulk bands (see Fig. 4b, Section SI III in online Supplementary Information).

Topological surface states are completely determined by the bulk band-structure topology \cite{TIbasic}, ensuring that superconducting symmetry breaking in the bulk bands must necessarily effect the surface spectrum. Three dimensional topological insulators doped into a superconducting phase are classified by an integer invariant (``n"), rather than a Z$_2$ invariant (parity invariant) \cite{FuNew}. The specific band structure and strong spin orbit coupling of Cu$_x$Bi$_2$Se$_3$ suggest that the topological invariant ``n" may be non-zero, allowing the possibility of topological superconductivity in Cu$_x$Bi$_2$Se$_3$. This is further supported by the fact that the ratio $\frac{m}{\mu}$ for Cu$_x$Bi$_2$Se$_3$ is well into the topological regime \cite{FuNew} as seen in our data. The exact nature of the superconducting order-parameter will depend on the parity of the pairing wavefunction, which may be even or odd under inversion since Bi$_2$Se$_3$ is a centrosymmetric crystal. Of the two remaining possibilities, even parity will result in a fully gapped band structure, and odd parity will generate new surface states resembling those shown in Fig. 4d \cite{FuNew}. In either scenario, the preserved topological spin polarized surface state will allow the superconducting wavefunction to host Majorana fermions \cite{FuSCproximity}, which might be manipulated adiabatically if the surface is fully gapped (even parity, Case 1 in Fig. 4). In the second, odd parity case, the material will be the first known realization of a state of matter specifically termed ``topological superconductivity" \cite{FuNew,SchnyderClass,KitaevClass}. While the exact determination of order-parameter phase will require low temperature phase sensitive measurements on ultra-high purity large single crystals, the unusual doping evolution towards achieving superconductivity discovered here provides critical clues for understanding pairing in doped topological insulators, as well as constituting the key ingredients for developing a general theory of these novel Bi-based superconductors.

\textbf{Methods summary:}

Experimental observation of surface and bulk separation and their microscopic electron kinetics constitutes overcoming several significant technical challenges. (see online Supplementary Information)

Angle resolved photoemission spectroscopy (ARPES) measurements were performed at the Advanced Light Source beamlines 10 and 12 using 35.5-48 eV photons and Stanford Synchrotron Radiation Laboratory (7-22eV photons) with better than 15 meV energy resolution and overall angular resolution better than 1$\%$ of the Brillouin zone (BZ). Samples were cleaved and measured at 15$^o$K, in a vacuum maintained below 8$\times$10$^{-11}$ Torr. Momentum along the $\hat{z}$ axis is determined using an inner potential of 9.5 eV, consistent with previous photoemission investigations of undoped Bi$_2$Se$_3$ \cite{MatthewNatPhys}. Large single crystals of Cu$_x$Bi$_2$Se$_3$ were grown using methods described in the supplementary information. Surface and bulk state band calculations were performed for comparison with the experimental data, using the LAPW method implemented in the WIEN2K package \cite{wien2k}. Details of the calculation are identical to those described in ref. \cite{MatthewNatPhys}.

\textbf{Corresponding author:}

Correspondence and requests for materials should be addressed to M.Z.H. (mzhasan@Princeton.edu).

\textbf{Acknowledgements:}

We acknowledge helpful discussions with L. Fu, A. Kitaev, A. Ludwig and F.D.M. Haldane. We are grateful for beamline support from S.-K. Mo, M. Hashimoto, D.-H. Lu and R. Moore.

\newpage

\begin{figure*}[t]
\includegraphics[width = 11cm]{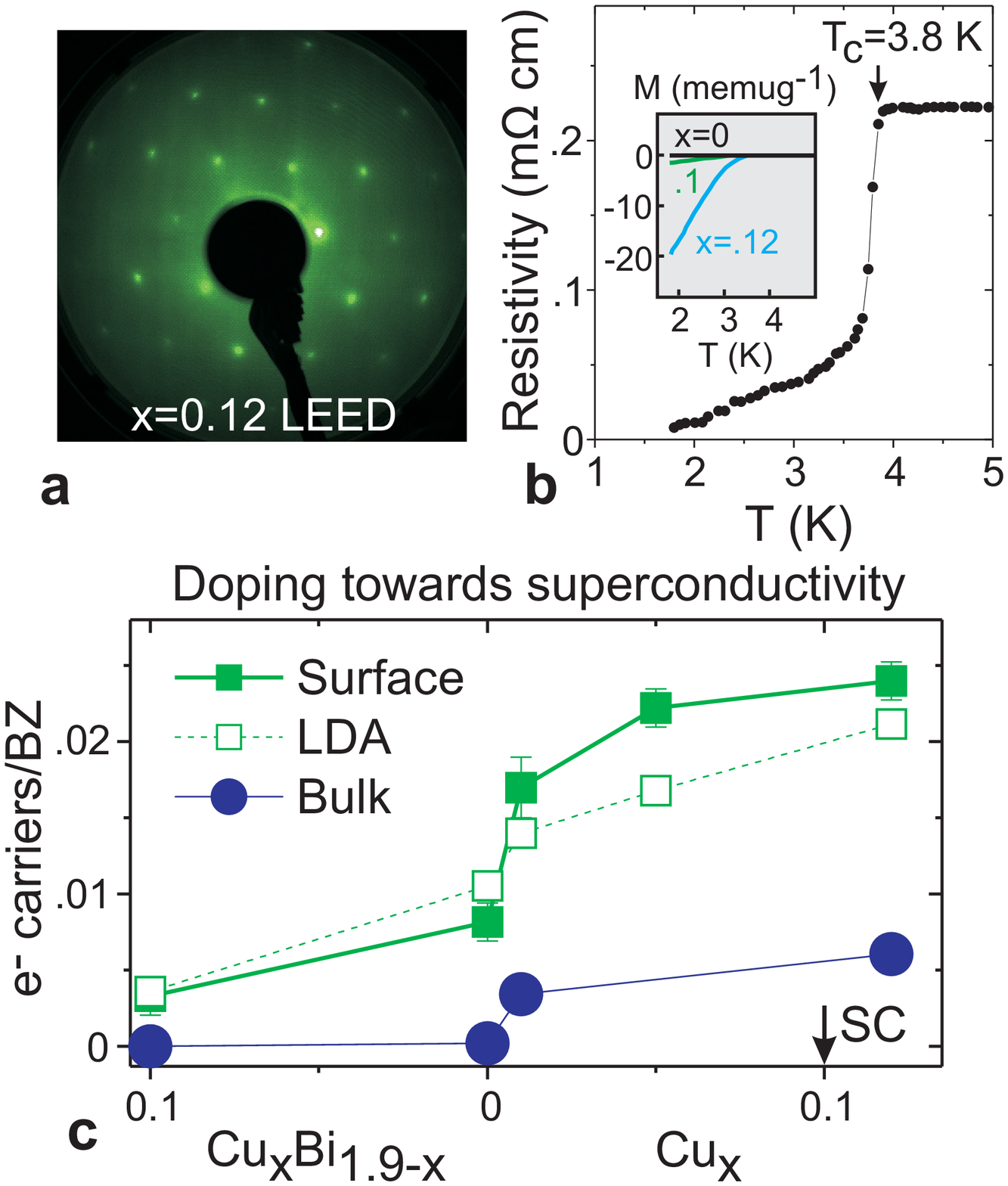}
\caption{{\bf{Superconductivity in Cu$_x$Bi$_2$Se$_3$ crystals}}: \textbf{a}, A low-energy electron diffraction image taken at 200eV electron energy provides evidence for a well ordered surface with no sign of superstructure modulation. \textbf{b}, Resistivity and magnetic susceptibility measurements for samples used in this study. (detailed materials characterization is reported in Ref.\cite{HorSC}) Samples exhibit a superconducting transition at 3.8$^o$K at optimal copper doping (x=0.12). \textbf{c}, The number of charge carriers is calculated from the Luttinger count ($\frac{FS\,area}{BZ\,area}$, $\times$2 for the doubly degenerate bulk band). LDA predictions show the carrier density obtained by aligning LDA band structure with the experimentally determined binding energy of the Dirac point.}
\end{figure*}

\begin{figure*}[t]
\includegraphics[width = 13cm]{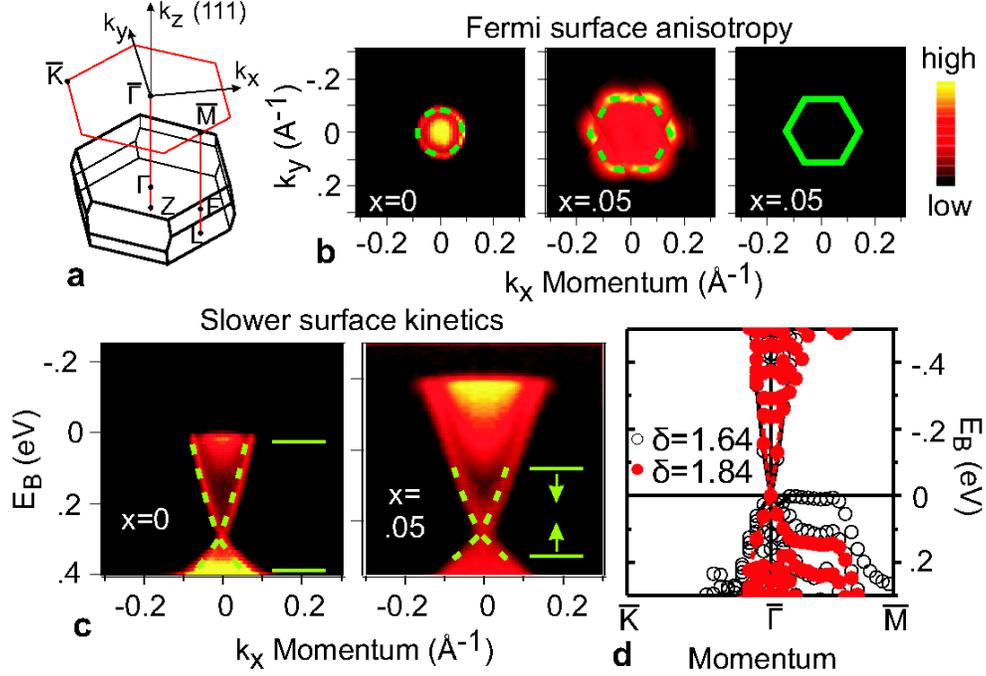}
\caption{{\bf{Surface electron kinematics of Cu$_x$Bi$_2$Se$_3$}}: \textbf{a}, The hexagonal surface Brillouin zone of Cu$_x$Bi$_2$Se$_3$ is shown above a diagram of the three dimensional bulk Brillouin zone. \textbf{b}, Symmetrized surface state Fermi surfaces are displayed with (right) a GGA prediction based on the experimental Fermi level. \textbf{c}, Photoemission measurements are shown at high photon energy (E$>$20 eV) for non-superconducting Cu$_x$Bi$_2$Se$_3$, demonstrating a reduced surface state dispersion after copper is added. \textbf{d}, When the Bi-Se plane spacing at the surface of a 12-layer slab is increased by 0.2 $\AA$ (1.64 to 1.84 $\AA$), dispersion in the upper Dirac cone increases by 16$\%$.}
\end{figure*}

\begin{figure*}[t]
\includegraphics[width = 16cm]{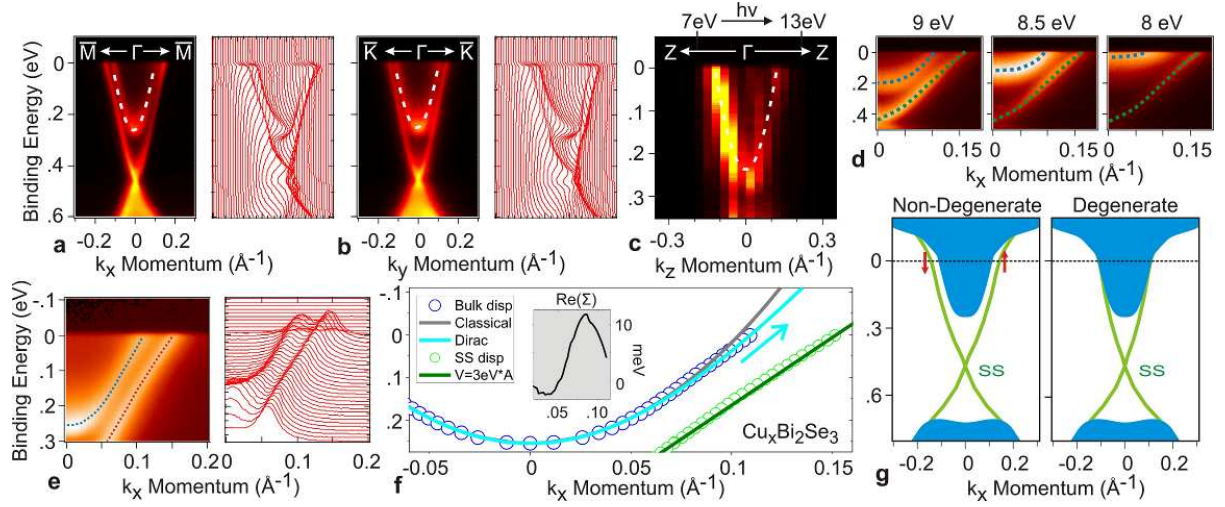}
\caption{{\bf{Band topology at the superconducting composition}}: \textbf{a}-\textbf{b}, Momentum dependence of the bulk and surface conduction bands in superconducting Cu$_{0.12}$Bi$_2$Se$_3$ is measured with low energy (9.75eV) photons for enhanced bulk sensitivity. \textbf{c}, Dispersion along the z-axis is examined by varying incident photon energy. \textbf{d}, Bulk and surface bands remain separate at intermediate k$_z$ values. \textbf{e}-\textbf{f}, Energy of the bulk electrons is compared with Dirac (v$_C$=6 eV$\cdot$$\AA$) and classical (parabolic) fits with a mass of 0.155M$_e$. An inset shows self-energy with respect to the Dirac fit. \textbf{g}, The surface electronic structure presents a non-trivial topological setting for superconductivity because (green) surface and (blue) bulk bands do not overlap.}
\end{figure*}

\begin{figure*}[t]
\includegraphics[width = 13cm]{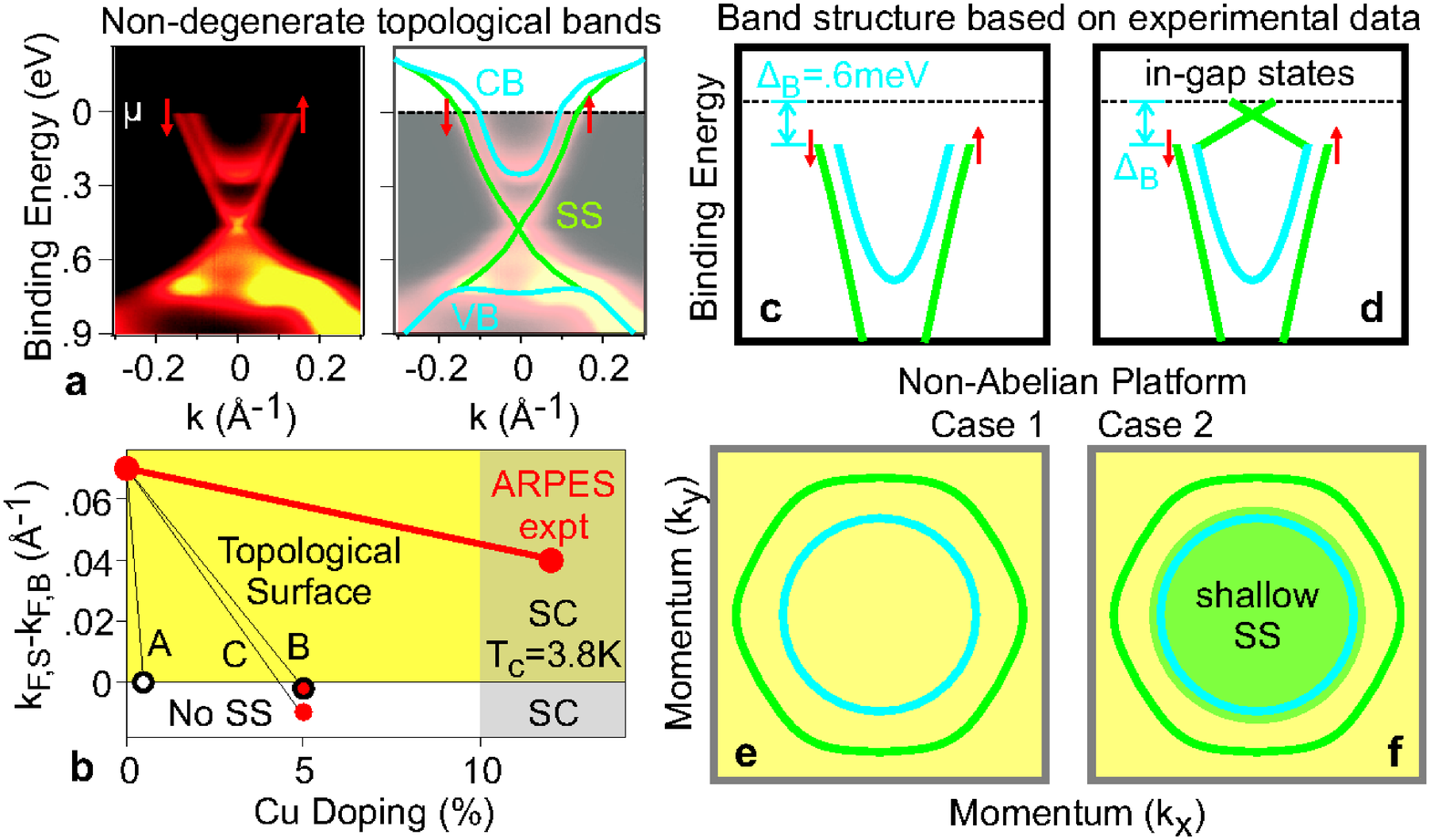}
\caption{{\bf{Symmetry breaking on the topological surface:}} \textbf{a}, Bulk and surface electrons are non-degenerate at the Fermi level. \textbf{b}, A phase diagram compares (\textbf{expt}) the measured superconducting topology with preliminary expectations based on cases in which (\textbf{A}) each Cu atom donates one doped electron, (\textbf{B}) the experimental chemical potential is applied to GGA band structure, and (\textbf{C}) the experimental chemical potential is applied to the band structure of undoped Bi$_2$Se$_3$. Electronic states expected below T$_C$ are drawn for (\textbf{c},\textbf{e}) even parity superconductivity and (\textbf{d},\textbf{f}) an example of odd parity ``topological superconductivity". Panels \textbf{e}-\textbf{f} show states within 5 meV of the Fermi level.}
\end{figure*}


\begin{thebibliography}[

\bibitem{Intro} Moore, J. E. Topological insulators: The next generation. \emph{Nature Physics} \textbf{5}, 378 (2009).
    http://www.nature.com/nphys/journal/v5/n6/full/nphys1274.html
\bibitem{TI_RMP} Hasan, M. Z. $\&$ Kane, C. L. Colloquium: Topological Insulators. \emph{Rev. Mod. Phys} \textbf{82}, 3045 (2010); Preprint at http://arxiv.org/abs/1002.3895 (2010); X.-L. Qi and S.-C. Zhang, Preprint at http://arxiv.org/abs/1008.2026 (2010).

\bibitem{TIbasic} Fu, L. $\&$ Kane, C. L. Topological insulators with inversion symmetry. \emph{Phys. Rev. B} \textbf{76}, 045302 (2007).
\bibitem{DavidNat1} Hsieh, D. \emph{et al.} A topological Dirac insulator in a quantum spin Hall phase. \emph{Nature} \textbf{452}, 970-974 (2008) [Submitted in \textbf{2007}].
\bibitem{DavidScience} Hsieh, D. \emph{et al.} Observation of unconventional quantum spin textures in topological insulators. \emph{Science} \textbf{323}, 919-922 (2009).
\bibitem{DavidTunable} Hsieh, D. \emph{et al.} A tunable topological insulator in the spin helical Dirac transport regime. \emph{Nature (London)} \textbf{460}, 1101-1105 (2009).
\bibitem{MatthewNatPhys} Xia, Y. \emph{et al.} Observation of a large-gap (3D) topological-insulator class with a single Dirac cone on the surface. \emph{Nature Physics} \textbf{5}, 398-402 (2009). Preprint at http://arxiv.org/abs/0812.2078 (2008).
\bibitem{ChenBiTe} Chen, Y.L. \emph{et al.}, Experimental Realization of a Three Dimensional Topological Insulator. \emph{Science} \textbf{325}, 178-181 (2009).
\bibitem{PairStr} Bray-Ali, N., Ding, L. $\&$ Haas, S. Topological order in paired states of fermions in two-dimensions with breaking of parity and time-reversal symmetries. \emph{Phys. Rev. B} \textbf{80}, 180504(R) (2009).
\bibitem{MudrySurfSC} Santos, L. \emph{et al.} Superconductivity on the surface of topological insulators and in two-dimensional noncentrosymmetric materials. \emph{Phys. Rev. B} \textbf{81}, 184502 (2010).
\bibitem{NagaosaUnconventional} Linder, J. \emph{et al.} Unconventional superconductivity on a topological insulator. \emph{Phys. Rev. Lett.} \textbf{104}, 067001 (2010).
\bibitem{FuNew} Fu, L. $\&$ Berg, E. Odd-parity topological superconductors: theory and application to Cu$_x$Bi$_2$Se$_3$. Preprint at $\langle$http://arxiv.org/abs/0912.3294$\rangle$ (2009).
\bibitem{QiTopoSC} Qi, X.-L., Hughes, T. L. $\&$ Zhang, S.-C. Fermi surface topological invariants for time reversal invariant superconductors. \emph{Phys. Rev. B} \textbf{81}, 134508 (2010).
\bibitem{AshvinChiral} Hosur, P., Ryu, S. $\&$ Vishwanath, A. `Chiral' topological insulators, superconductors and other competing orders in three dimensions. \emph{Phys. Rev. B} \textbf{81}, 045120 (2010).
\bibitem{ImpurityStates} Biswas, R. R. $\&$ Balatsky, A. V. Impurity-induced states on the surface of 3D topological insulators. \emph{Phys Rev B} \textbf{81}, 233405 (2010).
\bibitem{FerroSplitting} Garate, I. $\&$ Franz, M. Inverse spin-galvanic effect in a topological-insulator/ferromagnet interface. \emph{Phys. Rev. Lett.} \textbf{104}, 146802 (2010).
\bibitem{FuHexagonal} Fu, L. Hexagonal warping effects in the surface states of topological insulator Bi$_2$Te$_3$. \emph{Phys. Rev. Lett.} \textbf{103}, 266801 (2009).
\bibitem{SchnyderClass} Schnyder A. P. \emph{et al.} Classification of topological insulators and superconductors in three spatial dimensions. \emph{Phys. Rev. B} \textbf{78}, 195125 (2008).
\bibitem{KitaevClass} Kitaev, A. Periodic table for topological insulators and superconductors. \emph{Proceedings of the L.D. Landau Memorial Conference ``Advances in Theoretical Physics"}, (June 22-26, 2008).
\bibitem{FuSCproximity} Fu, L. $\&$ Kane, C. L. Superconducting proximity effect and Majorana Fermions at the surface of a topological insulator. \emph{Phys. Rev. Lett.} \textbf{100}, 096407 (2008).
\bibitem{HorSC} Hor, Y. S. \emph{et al.} Superconductivity in Cu$_x$Bi$_2$Se$_3$ and its implications for pairing in the undoped topological insulator. \emph{Phys. Rev. Lett.} \textbf{104}, 057001 (2010).
\bibitem{HorAllSC} Hor, Y. S. \emph{et al.} Superconductivity and non-metallicity induced by doping the topological insulators Bi$_2$Se$_3$ and Bi$_2$Te$_3$. Preprint at $\langle$http://arxiv.org/abs/1006.0317$\rangle$ (2010).
\bibitem{BiTeSbTe} Hsieh, D. \emph{et al.} Observation of time-reversal-protected single-Dirac-cone topological-insulator states in Bi$_2$Te$_3$ and Sb$_2$Te$_3$. \emph{Phys. Rev. Lett.} \textbf{103}, 146401 (2009).
\bibitem{CuAmphoteric} Vasko, A. Amphoteric nature of copper impurities in Bi$_2$Se$_3$ crystals. \emph{Appl. Phys.} \textbf{5}, 217-221 (1974).
\bibitem{SCproperties} Carlson, E. W. \emph{et al.} in \emph{The Physics of Conventional and Unconventional Superconductors}, edited by Bennemann, K. H. $\&$ Ketterson, J. B. (Springer-Verlag, Berlin, 2002).
\bibitem{SCDong} Qian, D. \emph{et al.} Low-lying quasiparticle states and hidden collective charge instabilities
in parent cobaltate superconductors. \emph{Phys. Rev. Lett.} \textbf{96}, 216405 (2006).
\bibitem{wien2k} Blaha, P. \emph{et al. Computer Code WIEN2K} (Vienna University of Technology, Vienna, 2001).



\end{thebibliography}
\end{document}